\RequirePackage{fix-cm}
\documentclass[smallcondensed]{svjour3}     
\smartqed  
\usepackage{graphicx}
\usepackage{siunitx}

\begin{document}

\title{Next Generation Muon g-2 Experiment at FNAL 
}


\author{Martin Fertl on behalf of the FNAL E989 collaboration}


\institute{M. Fertl \at
				Center for Experimental Nuclear Physics and Astrophysics\\
              University of Washington \\
              Tel.: +1-206-543-9134\\
              \email{mfertl@uw.edu}        
}

\date{Received: date / Accepted: date}

\maketitle

\begin{abstract}

The precise measurement of the muon anomalous magnetic moment $a_\mathrm{\mu}$ has stimulated much theoretical and experimental efforts over more than six decades. The last experiment at Brookhaven National Laboratory, Upton, NY, USA obtained a value more than three standard deviations larger than predicted by the Standard Model of particle physics, and is one of the strongest hints for physics beyond the Standard Model.
A new experiment at Fermi National Accelerator Laboratory, Batavia, IL, USA to measure $a_\mathrm{\mu}$ with fourfold increased precision to \SI{140}{ppb} is currently in its commissioning phase. While the new experiment will reuse the \SI{1.45}{\tesla} superconducting muon storage ring which was shipped from Brookhaven National Laboratory, most of the other instrumentation of the experiment will be new. This will allow the experiment to make efficient use of the significantly higher number of muons available at the new muon campus.
We discuss the general status of the experiment and in particular focus on the improved tools available to homogenize and determine the magnetic field in the muon storage ring.


\keywords{muon \and magnetic moment \and precision magnetic field \and nuclear magnetic resonance}
\PACS{06.30.Ft \and 07.55.Ge \and 14.60.Ef}
\end{abstract}

\section{Introduction}
\label{sec:1}

The measurement of the anomalous magnetic moment of the muon provides a particularly sensitive test to physics beyond the Standard Model of particle physics (SM) as it can be calculated $a_\mu^{\mathrm{SM}}$ and measured $a_\mu^{\mathrm{exp}}$ to high precision ($a_\mathrm{\mu} = \left(g-2\right)/2$ is the muon magnetic anomaly) \cite{LeptonDipole2010}. This comparison provides a completeness test of the SM with respect to new particles and interactions whose contributions would alter the theoretical prediction of $a_\mu^{\mathrm{SM}}$.
The last result obtained by the E821 experiment at Brookhaven National Laboratory (BNL) deviates from the most accurate SM predictions by
\begin{equation}
\Delta a_\mu = a_\mu^{\mathrm{exp}}-a_{\mu}^{\mathrm{SM}} = \left[\left(263-289 \right)\pm 80 \right] \cdot 10^{-11},
\end{equation}
which corresponds to a more than three standard deviation (3$\sigma$) discrepancy. The interval represents uncertainties related to different calculations of the so called hadronic vacuum polarization (HVP) contribution \cite{Davier2011,Teubner2012} and the hadronic light by light scattering (HLbL) contribution to $a_\mu^{\mathrm{SM}}$.

At this symposium the status of two new experiments to measure the muon anomalous magnetic moment $a_\mathrm{\mu}$ were presented. 
The approach taken by the J-PARC E34 experiment is described in \cite{Mibe2015,Marshall2015}. Here we concentrate on the E989 experiment at Fermi National Accelerator Laboratory (FNAL) which has passed a very important milestone since the symposium: the successful cool down and ramping of the superconducting storage ring to its nominal field for the first time since 2001. This provided the starting point for a several-month-long effort to shim the magnet to the desired field homogeneity.

\section{Experimental techniques}
\label{sec:2}

The anomalous magnetic moment $a_\mu$ gives rise to a small difference between the muon spin precession frequency $\omega_\mathrm{s}$ and the muon cyclotron frequency $\omega_\mathrm{c}$ for muons in a magnetic storage ring. The so called anomalous spin precession frequency $\omega_\mathrm{a} = \omega_\mathrm{s} -\omega_\mathrm{c}$. This frequency difference can be directly measured for muons stored in a magnetic field where they perform cyclotron motion with an angular frequency given by
\begin{equation}
\omega_\mathrm{c} = \frac{e B}{m c \gamma},
\end{equation}
where $e$ ($m$) is the muon's electric charge (rest mass), $B$ the local magnetic flux density, $c$ the speed of light and $\gamma$ the Lorentz factor.
The muon spin precession rate is given by
\begin{equation}
\omega_\mathrm{s} = \frac{g e B}{2 m c} + \left(1-\gamma\right) \frac{e B}{m c \gamma},
\end{equation}
where $g$ is the gyromagnetic factor of the muon.
The instantaneous anomalous spin precession frequency is then given by
\begin{equation}
\omega_\mathrm{a} = \frac{g-2}{2} \frac{e B}{m c} \equiv a_\mu \frac{e B}{m c}.
\end{equation}
A precise measurement of $\omega_\mathrm{a}$ and of the magnetic field allows to obtain $a_\mu$. Experimentally $a_\mu$ is extracted as the ratio
\begin{equation}
a_\mu =\frac{\frac{\omega_a}{\tilde{\omega}_p}}{\frac{\mu_\mathrm{p}}{\mu_\mathrm{\mu}}-\frac{\omega_a}{\tilde{\omega}_p}},
\end{equation}
where $\tilde{\omega}_\mathrm{p}$ is the equivalent Larmor frequency of a free proton averaged over the muon storage volume and $\mu_\mathrm{p}$/$\mu_\mathrm{\mu}$ is the ratio of the magnetic moments of free protons and muons as determined in the muonium hyperfine splitting experiment \cite{Liu1999}.

In the idealized case of a pure magnetic field $\omega_\mathrm{a}$ is independent of the muon kinetic energy. 
But to confine muons with non zero vertical momentum to the toroidal storage region of the storage ring, vertical focusing of the muon beam with electrostatic quadrupoles is employed. In this situation the anomalous spin precession vector is given by
\begin{equation}
\vec{\omega}_\mathrm{a} = - \frac{e}{m}\left[a_\mathrm{\mu} \vec{B}  - \left(a_\mathrm{\mu}-\frac{1}{\gamma^2-1}\right)\frac{\vec{\beta} \times \vec{E}}{c}\right],
\end{equation}
where $\vec{E}$ and $\vec{\beta}$ are the local electric field and muon velocity vectors. Choosing the so called muon magic momentum, $p_\mathrm{\mu}=\SI{3.094}{\giga\electronvolt}$, is equivalent to $\gamma =\num{29.3}$ which eliminates the electric field dependent contribution to $\vec{\omega_\mathrm{a}}$. Small systematic corrections have to be applied to account for motional magnetic field effects which arise from $\vec{\beta} \times \vec{E} \neq \vec{0}$ and $\gamma\neq 29.3$ \cite{Bennett2006}. For magic momentum muons in a \SI{1.45}{\tesla} magnetic field the cyclotron period is \SI{149}{\nano\second}.  

The FNAL E989 experiment uses the concept of a self analyzing polarimeter to measure $\omega_\mathrm{a}$.
Charged pions emitted in the forward direction of a high intensity pion production target are captured in a transport beam line where they decay in flight ($\pi^+ \rightarrow \mu^+ + \nu_\mu $) and build up a high intensity, highly polarized muon beam. The muon beam is injected into the muon storage ring and a pulsed kicker magnet transports the muon beam pulse onto a stable storage orbit. For $g>2$ the anomalous spin precession causes the muon spin to advance the muon momentum vector as the muons circulate around the ring. The parity violating decay $\mu^+ \rightarrow e^+ + \bar{\nu}_\mathrm{\mu} + \nu_\mathrm{e}$ has a differential decay rate in the muon rest frame proportional to $\left(1+ A\left(E\right) \cos \theta \right)$ where $\theta$ is the angle between muon spin and decay positron momentum vector and $A\left(E\right)$ is a positron energy dependent asymmetry factor. Positrons are emitted preferentially along the muon spin direction. In the case of $\theta=0$ the positrons are Lorentz boosted to higher energies in the laboratory frame while for $\theta=\pi$ the Lorentz boost reduces the positron energy in the laboratory. Thus a calorimeter station which detects the inwards spiraling positrons will see a count rate modulation of higher and lower energy particles with the anomalous spin precession frequency $\omega_\mathrm{a}$. The calorimeter station records the positron hit time (relative to the muon beam injection time) and energy. The registered events are then filled in a histogram as a function of calorimeter hit time. Requiring a positron kinetic energy $>\SI{1.8}{\giga\electronvolt}$ optimizes the statistical sensitivity across all positron energy groups.

To obtain $\tilde{\omega}_p$ the magnetic field in the muon storage region is measured using pulsed nuclear magnetic resonance (pNMR) probes. A trolley equipped with 17
pNMR probes distributed across the muon storage region is employed to measure (approximately once per day) the magnetic field distribution around the storage ring. In this time no muons can be stored in the ring. During muon storage operation the time stability of the magnetic field is surveyed with an array of about 380 pNMR probes embedded in the walls of the muon storage ring vacuum chamber. To express the magnetic field measured with the pNMR probes in terms of a hypothetical measurement with free protons a cross calibration scheme is under development.

\section{Improved muon source and particle detectors}
\label{sec:3}
The FNAL E989 collaboration will use the new FNAL muon campus infrastructure to accumulate 21 more decay positrons than observed in the BNL E821 experiment. The total uncertainty budget of \SI{140}{ppb} for the FNAL E989 experiment is split into \SI{100}{ppb} and \SI{70}{ppb} for the statistical and systematic uncertainties on $\omega_\mathrm{a}$ and \SI{70}{ppb} for the overall uncertainty of the proton precession frequency $\tilde{\omega}_\mathrm{p}$ \cite{TDR2015}. 

The upgraded linear accelerator and booster ring structure of FNAL will deliver proton pulses (\SI{8}{\giga\electronvolt}, \num{4E12} $p$ per pulse, \SI{1.3}{\second} pulse separation) which will be re-bunched into four pulses of \num{1E12} $p$ and \SI{121}{\nano\second} pulse length in the recycler ring before impinging on an Inconel 600 $\pi^+$ production target with \SI{10}{\milli\second} time separation. The secondary $\pi^+$ beam will be focused with a pulsed lithium lens into the transport beam line which accepts $\pi^+$ in the forward direction with a momentum spread of $\pm\SI{0.5}{\percent}$ around \SI{3.11}{\giga\electronvolt\per c}. In the transport beam line and in the delivery ring section the in-flight-decay of $\pi^+$ generates the polarized $\mu^+$ beam. The nearly eight times longer flight distance at FNAL compared to BNL allows the residual hadronic contamination in the muon beam to decay away before it reaches the muon storage ring. This will suppress substantially the so called hadronic flash in the positron calorimeters after muon beam injection which was a major source of background for the BNL experiment.
The muons are injected into the storage ring through a hole in the back yoke of the storage ring magnet. A detailed Monte-Carlo study of the muon beam dynamics in the storage ring is performed to optimize the number of stored muons which will lead to an increased instantaneous count rate of decay positrons. Therefore a new, high count rate capable, and highly segmented electromagnetic calorimeter will be employed. Twenty-four individual calorimeter stations, each consisting of an array of $6 \times 9$ $\mathrm{PbF_2}$ crystals ($\SI{25.4}{\milli\meter}\times\SI{25.4}{\milli\meter} \times \SI{152.4}{\milli\meter}$) will be spaced equidistantly around the inner radius of the storage ring. Each crystal is individually instrumented with a silicon photomultiplier (SiPM) to detect the Cerenkov light generated by the high energy positrons. The high segmentation allows hit position discrimination while the fast SiPM response can separate events as close as \SI{3}{\nano\second} (\SI{800}{\mega\hertz} digitization rate) \cite{Fienberg2015} which will allow to address pile-up related systematic effects. In addition SiPMs, unlike traditional photo multiplier tubes, can be employed in the stray field of the magnet and do not distort the pristine magnetic field of the storage ring. Therefore SiPMs (incl. the preamplifier electronics) can be directly glued onto the $PbF_2$ crystals, eliminating the need for a long light guide which in turn improves the energy resolution of the calorimeter due to the larger number of detected photons per positron passing the crystal.
Retractable fiber harp detectors will be installed in the muon storage region to measure the muon distribution in the storage region. Straw tracker stations will be operated in front of three positron calorimeters which will allow the precise reconstruction of the positron flight path.

\section{Precision magnetic field}
\label{sec:PrecMagField}
\begin{figure}
\begin{center}
\includegraphics[width=0.45\textwidth]{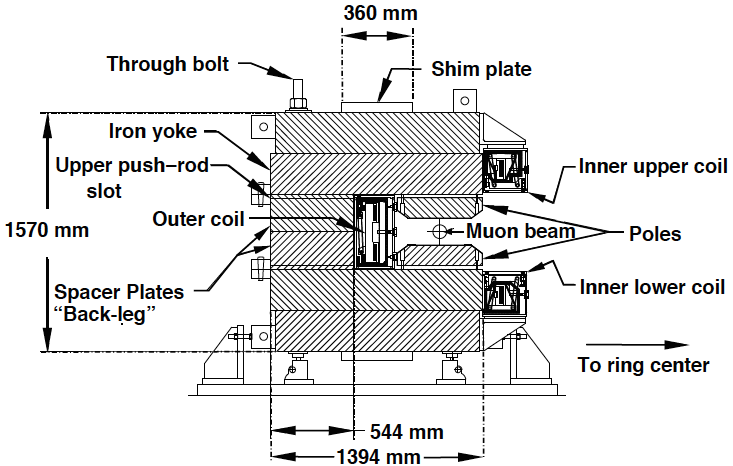}\includegraphics[width=0.45\textwidth]{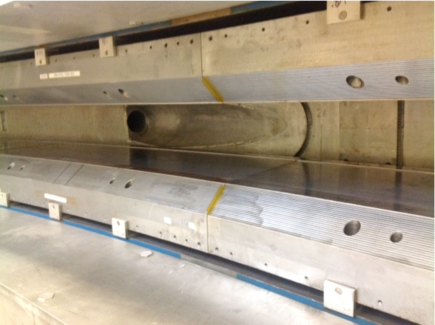}
\caption{Section through the C-shaped super-ferric ring magnet and picture of the inflector port between the pole pieces.
Reprinted with permission from \cite{Bennett2006}.}
\label{fig:1}
\end{center}
\end{figure}

\begin{figure}
\begin{center}
\includegraphics[width=0.6\textwidth]{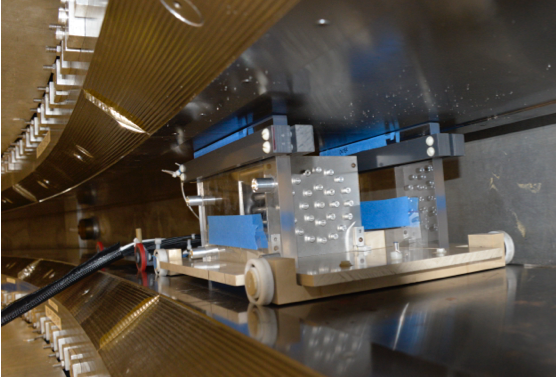}
\caption{A subset of 25 pNMR probes installed in a circular matrix and mounted to the shimming cart. The pNMR probes cover the area of the muon storage region. Magnetic field measurements are taken at about 6000 azimuthal positions around the storage ring.}
\label{fig:2}
\end{center}
\end{figure}

In September 2015 the superconducting magnet was ramped to its nominal field of \SI{1.45}{\tesla} for the first time since 2001. This started a several-months-long campaign to homogenize the magnetic field around the muon storage region. To achieve the anticipated \SI{70}{ppb} uncertainty on the magnetic field average across the toroidal muon storage region (\SI{9}{\centi\meter} minor and \SI{7}{\meter} major diameter) passive and active shimming techniques and high precision measurement instrumentation have been developed. 
Figure \ref{fig:1} shows a section through the C-shaped yoke of the super-ferric muon storage ring which consists of \SI{30}{\degree} wide, cast iron pieces surrounding the four super conducting coils. In each yoke section three ten-degree-wide pole pieces are installed at the top and bottom, leaving enough room to adjust the radial position of wedge-shaped iron pieces. The exact dimension of the air gaps are used to mainly adjust the dipole strength of the magnetic field while the exact positions and orientations of the pole pieces and wedge shims control the magnetic quadrupole moment of the field. The collaboration aims to achieve a \SI{10}{ppm} magnetic field homogeneity adjusting these passive shim elements.
Surface current coils, installed on the pole pieces, will be used to further improve the magnetic field homogeneity but will also allow to introduce controlled magnetic field gradients for systematic studies.
To measure the magnetic field about 400 improved pulsed proton nuclear magnetic resonance (pNMR) probes were built based on the initial design used in the BNL E821 experiment \cite{Prigl1996}. The improvements concern the mechanical as well as the electrical stability of the probes. About 380 pNMR probes will be installed in the wall of the muon storage ring vacuum chamber to survey the stability of the magnetic field in time. In addition a set of 17 pNMR probes will be installed in a circular matrix (radii \SI{0}{\milli\meter}, \SI{17.5}{\milli\meter}, and \SI{35}{\milli\meter}) on the so called \textsl{trolley}. This device can be automatically pulled through the muon storage region to measure the magnetic field distribution at about 6000 azimuthal positions to obtain a detailed map of the magnetic field around the ring. A multipole decomposition of the obtained field map is subsequently used to calculate the average magnetic field in the muon storage region.
During the initial shimming of the magnet a subset of \num{25} pNMR probes is installed in a circular matrix (radii \SI{0}{\milli\meter}, \SI{22.4}{\milli\meter}, and \SI{44.7}{\milli\meter}, see Figure \ref{fig:2}) and mounted on a cart riding between the pole pieces. A semi-automatic pulley system is used to take magnetic field measurements at about 6000 azimuthal positions around the ring. The obtained magnetic field distribution is subsequently decomposed in multipole moments and these measurements guide the decision of how to adjust the passive shimming elements to reduce the magnetic field homogeneity.

\section{Summary}
\label{sec:Summary}
We have reported the current status of the E989 experiment at FNAL, one of the two upcoming experiments aiming to improve on the last measurement of the muon anomalous magnetic moment by the BNL E821 experiment. We have discussed the new muon campus beamline infrastructure at FNAL which will provide the necessary number of muons needed to achieve a fourfold reduction of the experimental uncertainty. Furthermore we have presented the state-of-the-art particle detection systems that will be employed to handle the higher instantaneous count rate and to provide more information about beam dynamics, indispensable to control systematic effects on the extraction of the anomalous muon spin precession frequency. Improved instrumentation to measure and adjust the magnetic field in the muon storage region was discussed. 

%

\begin{acknowledgements}
The author thanks the organizers of this marvelous symposium and
his many colleagues on the Fermilab E989 g-2
experiment. The author is supported by the DOE, Office of Science, Office of Nuclear
Physics, award DE-FG02-97ER41020.
\end{acknowledgements}

\bibliographystyle{spphys}       
\bibliography{Fertl_Bibliography}   

\end{document}